\documentclass[aps,prl,twocolumn]{revtex4}
\usepackage{amsmath}
\usepackage{amssymb}
\usepackage{graphics}

\begin{document}
\title{Note on Possible Emergence Time of Newtonian Gravity}   
\author{Lajos Di\'osi}
%\email{diosi.lajos@wigner.mta.hu}
%\homepage{www.rmki.kfki.hu/~diosi}
\affiliation{
Wigner Research Center for Physics\\
H-1525 Budapest 114, POB 49, Hungary}
%\date{\today}

\begin{abstract}
If gravity were an emergent phenomenon, some relativistic as well as 
non-relativistic speculations claim it is, then a certain emergence time 
scale $\tau_?$ would characterize it. We argue that the available  
experimental evidences have poor time resolution regarding how immediate
the creation of Newton field of accelerated mass sources is. Although 
the concrete theoretical model of gravity's `laziness' is missing, the 
concept and the scale $\tau_?\sim1$ms, rooted in an extrapolation of 
spontaneous wave function collapse theory, might be tested directly 
in reachable experiments. \\~\\
tel/fax: +36-302956469/-13959151, e-mail: diosi.lajos@wigner.mta.hu
\end{abstract}

%\pacs{03.65.Yz, 42.50.Lc}

\maketitle

Physicists have always been speculating that gravity may emerge from a
structure of deeper level. Gravity has been resisting to relativistic 
quantum field theories despite their robust success all over the past 
fifty-sixty years. If gravity itself is not a relativistic quantized 
field, it might be induced by them. According to Sakharov \cite{Sak67}, 
the Casimir energy of quantized matter fields yields elastic forces 
on the background space-time, resulting in similar dynamics to Einstein's 
general relativity. This approach, its variants and refinements
(cf., e.g., \cite{Adl82,Vis02}) represent the main stream of `emergent gravity'
investigations. Alternative concepts \cite{Jac95,Ver11,Pad10} relate Einstein's theory 
to thermodynamics and derive the gravitational force from the entropy.   
Other speculations postpone the relativistic aspects, abandon quantum field theory,
but assume an intrinsic relationship between Newtonian (i.e., non-relativistic) 
gravity and quantum mechanics \cite{DioLuk87,Dio87,Jon95,Pen96,DeFMai02,Ges04,WezOosZaa08}. Some advocate 
phenomenological mechanisms for the emergence of the Newton interaction \cite{DeFMai02,DioPap09,Dio13}.

Our note restricts itself for emergence of the Newtonian, non-relativistic gravity.
If this emergence is real at all, it must be characterized by a certain \emph{emergence time} $\tau_?$
and the value of $\tau_?$ is expected to be longer than the typical time-scales of relativistic
emergence. It could depend on the wave length but we further simplify our assumption as to
look for a single time scale $\tau_?$. We mean that the gravitational field of an accelerated mass source
shall not immediately follow the Newton law but with a delay of about $\tau_?$.
Although the concrete model of emergence is missing, the option has been
discussed recently \cite{Dio13} as a plausible consequence of the gravity-related spontaneous 
wave function collapse theory \cite{Dio87,Pen96}.
The heuristic discussion concluded to the existence of a characteristic emergence time $\tau_?\sim1$ms.  

As we said, our proposal is purposely non-relativistic but its relationship to the Einstein theory 
must be stated. A slight laziness, like $\tau_?\sim1$ms, in creating gravity would not invalidate 
the Einstein theory for the large scale dynamics of space-time. The only available experimental 
(indirect) evidence of gravitational waves confirms the radiation of the Hulse-Taylor binary 
pulsar at period 7.75h \cite{Wag75}, no way would they be influenced by our proposal.
Further relativistic predictions of the Einstein theory are light deflection, lensing and delay in 
the presence of gravity: these effects have been confirmed (cf. \cite{CiuWhe95}) in the gravitational field 
of static or slowly moving sources of irrelevant acceleration to confront the above guess of $\tau_?$.

For the proposed short emergence time $\tau_?$, the peaceful coexistence with the above
gravitodynamic evidences is comforting. Yet, we propose a concrete modification of Newton gravity which is even more 
reassuring: it reduces to the standard theory for purely gravitational many-body systems. 
Perhaps no Galilean invariant many-body model is able to capture a finite emergence time. Nevertheless,
we know that standard field theories capture finite \emph{propagation times} hence we
tolerate the theoretical obstacles of finite \emph{emergence times}. They might be relaxed
in the framework of a future theory where certain `fields' ---dynamical and/or statistical---assist 
to massive bodies. For the time being, we propose a simplest phenomenology.

We start from the standard Newton law:
\begin{equation}\label{Phi}
\Phi(r,t)=\frac{-GM}{\vert r-x_t\vert}, 
\end{equation}
where $\Phi$ is the Newton potential at location $r$ and time $t$, created by the mass $M$ at
location $x_t$ at the same time $t$. We propose the following retarded-smoothened version of the 
standard Newton potential:   
\begin{equation}\label{Phi_tau}
\Phi(r,t)=\int_0^\infty\!\!\!\!\frac{-GM}{\vert r-x_{t-\tau}\vert}~~\mathrm{e}^{-\tau/\tau_?}d\tau/\tau_?, 
\end{equation}
but this cannot be the full proposal because the value of $\Phi$ depends on the choice of the
inertial frame. Suppose the source is free-falling in a certain, say, homogeneous
external gravitational field. And suppose that we use the same free-falling reference frame.
The equivalence principle, just the non-relativistic one, says that in the free-falling
reference frame physics goes as if we were in a gravity-free inertial frame. If, furthermore,
we use the co-moving frame where $\dot x_t\equiv0$ then our proposal (\ref{Phi_tau}) reduces to
the standard law (\ref{Phi}). This result is plausible: if the source is at rest in an inertial
(gravity-free) frame then its Newton potential is static, not delayed at all by the emergence time $\tau_?$.  
Therefore we require that our equation (\ref{Phi_tau}) be valid  
i) in the free-falling reference frame where $M\ddot x_t$ is equal to the non-gravitational
forces and ii) in the $t$-dependent co-moving system where $\dot x_t=0$. 

The second condition guarantees the Galilean boost-invariance of our proposal.       
The first condition (together with the second) guarantees that masses
performing inertal motion solely under graviational forces would
create the standard immediate Newton field, without the delay. Hence our modification
of the Newton law does not influence the planetary dynamics. It influences systems with
non-gravitational forces.   

Can we then find evidences pro or contra our proposal in accomplished laboratory experiments
on Newton theory? In a standard Cavendish experiment \cite{LutTow82}, 
a torsion balance measures the gravitational attraction produced by static source masses.
Because of static sources, time-resolution is beyond the scope of the standard Cavendish 
experiments. Fortunately, there are Cavendish experiments with moving sources. 
In the Gundlach-Merkowitz experiment the sources are revolving and a time-resolution
below $1$min seems available \cite{GunMer00}. The re-analysis of the experimental data would put an
upper limit on gravity's laziness $\tau_?$, stronger than ever. 
Furthermore, a precise measurement can be done 
at the gravity wave detectors \cite{VIRLIG}, too. While they cannot resolve the gravity wave 
propagation time from a moving nearby source (e.g.: a spinning dumbbell), they would perfectly 
resolve the emergence time in (and much below) the range of 1s. 

Let us apply our proposal to a laboratory source accelerated by non-gravitational
forces. Eq.~(\ref{Phi_tau}) has been postulated in the co-moving inertial frame where $\dot x_t=0$;
in the laboratory system it acquires the boost $-\dot x_t$:
\begin{equation}\label{Phi_tau_lab}
\Phi(r,t)=\int_0^\infty\!\!\!\!\frac{-GM}{\vert r-x_{t-\tau}-\dot x_t\tau\vert}~~\mathrm{e}^{-\tau/\tau_?}d\tau/\tau_?. 
\end{equation}
The lowest order expansion in $\tau_?$ yields
\begin{equation}\label{Phi_tau_lab_approx}
\Phi(r,t)=\frac{-GM}{\vert r-x_t\vert}\left(1+\frac{\ddot x_t^r \tau_?^2/2}{\vert r-x_t\vert}\right). 
\end{equation}
The correction of the Newton law is proportional to the radial acceleration $\ddot x_t^r$ of the
source. The emergent field $\Phi(r,t)$ is stronger/weaker if the source accelerates respectively toward/off 
the location $r$. If, e.g., the source is revolving at constant angular frequency $\Omega$ along a circle,
the field in the center of the orbit is enhanced by the factor $1+\Omega^2\tau_?^2/2$, valid for 
$\Omega\tau_?\ll1$.

Let's summarize our work. We noticed that the time-resolution of available experimental data would not
disclose a tiny temporal ``laziness'' $\tau_?$ of Newton gravity. We propose a delay time
of the order of $\tau_?\sim1$ms, coming from speculations on spontaneous wave function collapse. A minimalist modification
of the Newton law captures the delay $\tau_?$ in such a way that the dynamics of purely 
gravitational motion remains the standard one. Our proposal modifies the
Newton field of sources accelerated by non-gravitational forces that is typical in laboratory experiments.  
Even if the theoretical 
background of a possible emergence time $\tau_?$ is vague at the moment, reachable laboratory experiments should 
answer if $\tau_?$ can be that big as $1$ms. They would easily push the upper limit on $\tau_?$ much below $1$ms.  
Or, they might in principle find new physics with $\tau_?\sim1$ms (or with even bigger one), confirming or
at least encouraging the related quantum theoretical speculations \cite{Dio13}.  

Support by the Hungarian Scientific Research Fund under Grant No. 75129
and support by the EU COST Action MP100 are acknowledged. The author thanks Tam\'as Geszti and
P\'eter Kir\'aly for interesting discussions.


\begin{thebibliography}{99}
\bibitem{Sak67} A.D. Sakharov, Dokl. Akad. Nauk. SSSR {\bf 177}, 70 (1967); Gen. Rel. Grav. {\bf 32} 365 (2000). 
\bibitem{Adl82} S.L. Adler, Rev. Mod. Phys. {\bf 54}, 729 (1982). 
\bibitem{Vis02} M. Visser, Mod. Phys. Lett. A {\bf 17} 977 (2002).
\bibitem{Jac95} T. Jacobson, Phys. Rev. Lett. {\bf 75}, 1260 (1995).
\bibitem{Ver11} E. P. Verlinde, JHEP {\bf 04}, 029 (2011).
\bibitem{Pad10} T. Padmanabhan, Rep. Prog. Phys. {\bf 73} 6901 (2010). 
\bibitem{DioLuk87} L. Di\'osi and B. Luk\'acs, Annln. Phys. {\bf 44} 488 (1987);
              L.Di\'osi: {\it Quantum measurement and gravity for each other}, 
              p299 in: Quantum Chaos - Quantum Measurement, eds.: P.Cvitanovic, I.Percival and A. Wirzba 
              (Kluwer, Dordrecht, 1992).
\bibitem{Dio87} L. Di\'osi, Phys. Lett. {\bf 120A} 377 (1987); 
                L. Di\'osi, Phys. Rev. {\bf A40}, 1165 (1989).
\bibitem{Jon95} K.R.W. Jones, Aust. J. Phys. {\bf 48}, 1055 (1995).
\bibitem{Pen96} R. Penrose, Gen. Rel. Grav. {\bf 28}, 581 (1996);  
\bibitem{DeFMai02} S. De Filippo and F. Maimone, Phys. Rev. D {\bf 66} 044018 (2002);
              F. Maimone, G. Scelza, A. Naddeo and V. Pelino, Phys. Rev A {\bf 83} 062124 (2011).
\bibitem{Ges04} T. Geszti, Phys. Rev. {\bf A69}, 032110 (2004).
\bibitem{WezOosZaa08} J. Van Wezel, T. Oosterkamp, J. Zaanen, Phil. Mag. {\bf 88} 1005 (2008); 
              J. Van Wezel and J. Van der Brink, Phil. Mag. {\bf 88} 1659 (2008).
\bibitem{DioPap09} L. Di\'osi and T.N. Papp, Phys. Lett. {\bf 373A} 3244 (2009);
                  L. Di\'osi, J. Phys. Conf. Ser. {\bf 174} 012002 (2009).
\bibitem{Dio13} L. Di\'osi: {\it Gravity-related wave function collapse: Is superfluid He exceptional?},  
                arXiv:1302.5364; L. Di\'osi: {\it Gravity-related wave function collapse: mass density resolution},  
                arXiv:1302.5365. 
\bibitem{Wag75} R. V. Wagoner, Astrophys. J. Lett. {\bf 196}, L63 (1975).
\bibitem{CiuWhe95} I. Ciufuloni and J.A. Wheeler: {\it Gravitation and Inertia} (Princeton University Press, Princeton, 1995).   
\bibitem{LutTow82} G.G. Luther and G.R. Towler, Phys. Rev. Lett. {\bf 48}, 121 (1982).
\bibitem{GunMer00} J.H. Gundlach and S.M. Merkowitz, Phys. Rev. Lett. {\bf 85}, 2869 (2000).
\bibitem{VIRLIG} B.P. Abbott, et al. (LIGO Scientific Collaboration), Rep. Prog. Phys. {\bf 72}, 076901 (2009);
                 T. Accadia et al. (VIRGO Scientific Collaboration), J. Phys.: Conf. Ser. {\bf 203}, 012074 (2010).
\end{thebibliography}
\end{document}